\documentstyle[prb,aps,twocolumn,epsf]{revtex}

\begin{document}
\draft

\twocolumn[\hsize\textwidth\columnwidth\hsize\csname @twocolumnfalse\endcsname

\title
{\bf Diffusion of hydrogen in crystalline silicon}

\author{Sabrina B\'edard\cite{byline1} and Laurent J. Lewis\cite{byline2}}

\address{
D{\'e}partement de Physique et Groupe de Recherche en Physique et Technologie
des Couches Minces (GCM), Universit{\'e} de Montr{\'e}al, Case Postale 6128,
Succursale Centre-Ville, Montr{\'e}al, Qu{\'e}bec, Canada H3C 3J7
}

\date{\today}

\maketitle

\begin{center}
{\bf Submitted to {\em Physical Review B}}
\end{center}

\begin{abstract}

The coefficient of diffusion of hydrogen in crystalline silicon is calculated
using tight-binding molecular dynamics. Our results are in good quantitative
agreement with an earlier study by Panzarini and Colombo \protect{[}Phys.\
Rev.\ Lett.\ {\bf 73}, 1636 (1994)\protect{]}. However, while our
calculations indicate that long jumps dominate over single hops at high
temperatures, no abrupt change in the diffusion coefficient can be observed
with decreasing temperature. The (classical) Arrhenius diffusion parameters,
as a consequence, should extrapolate to low temperatures.

\end{abstract}

\pacs{PACS numbers: 66.30.Jt, 66.30.Dn}

\vskip2pc
]

\narrowtext

In spite of the tremendous efforts that have been engaged in determining the
rate of diffusion of hydrogen in crystalline silicon, a concensus on the
``true value'' has not yet been reached. Of course, the diffusion constant
varies strongly with temperature --- not necessarily in a perfect Arrhenius
manner --- making a precise determination of the diffusion parameters
difficult. Other complications arise from possible collective effects (as
opposed to tracer diffusion), low-temperature quantum effects, impurities and
defects, etc.

\begin{figure}
\vspace*{-0.05in}
\epsfxsize=3.0in \epsfbox{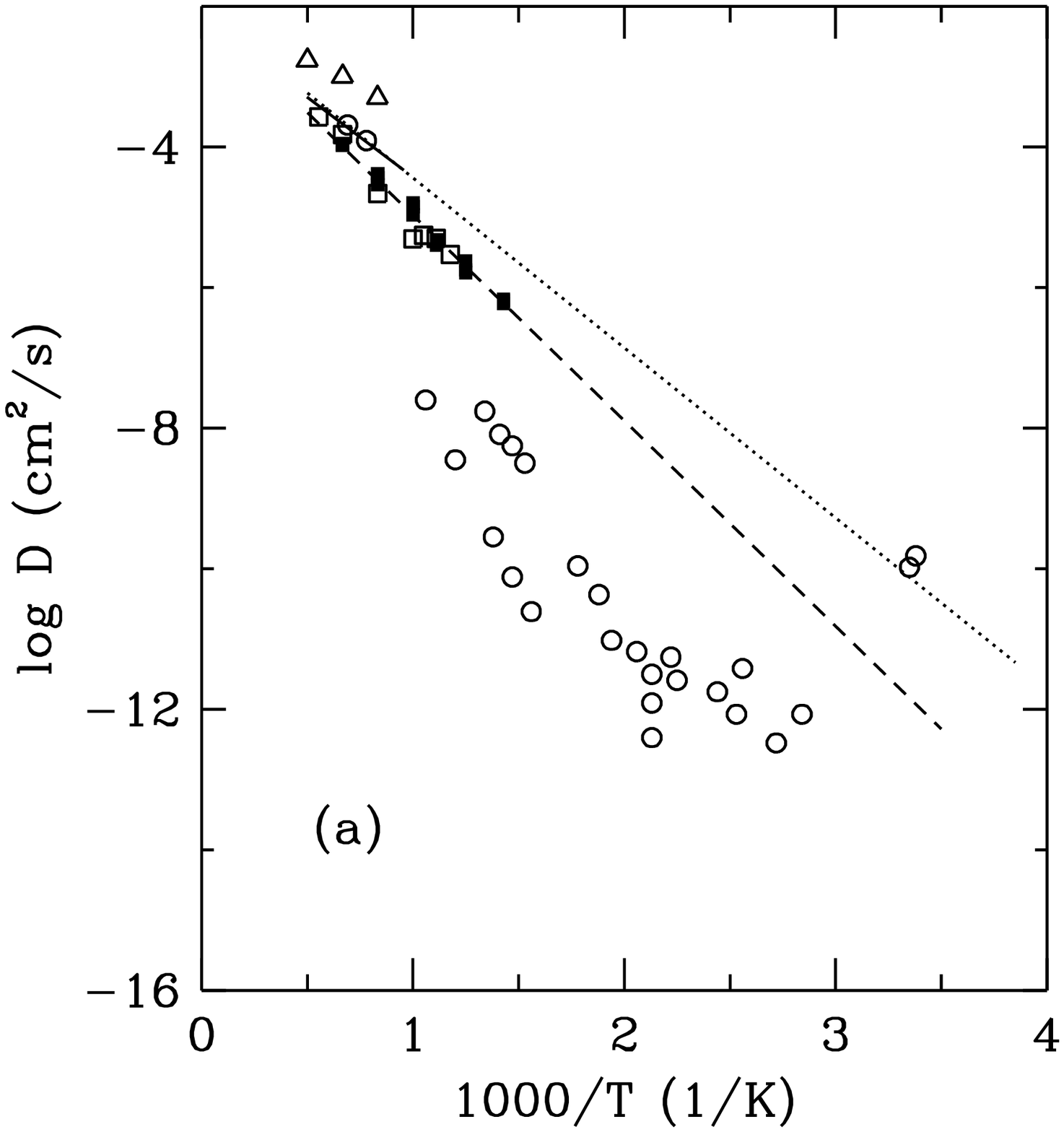}
\epsfxsize=3.0in \epsfbox{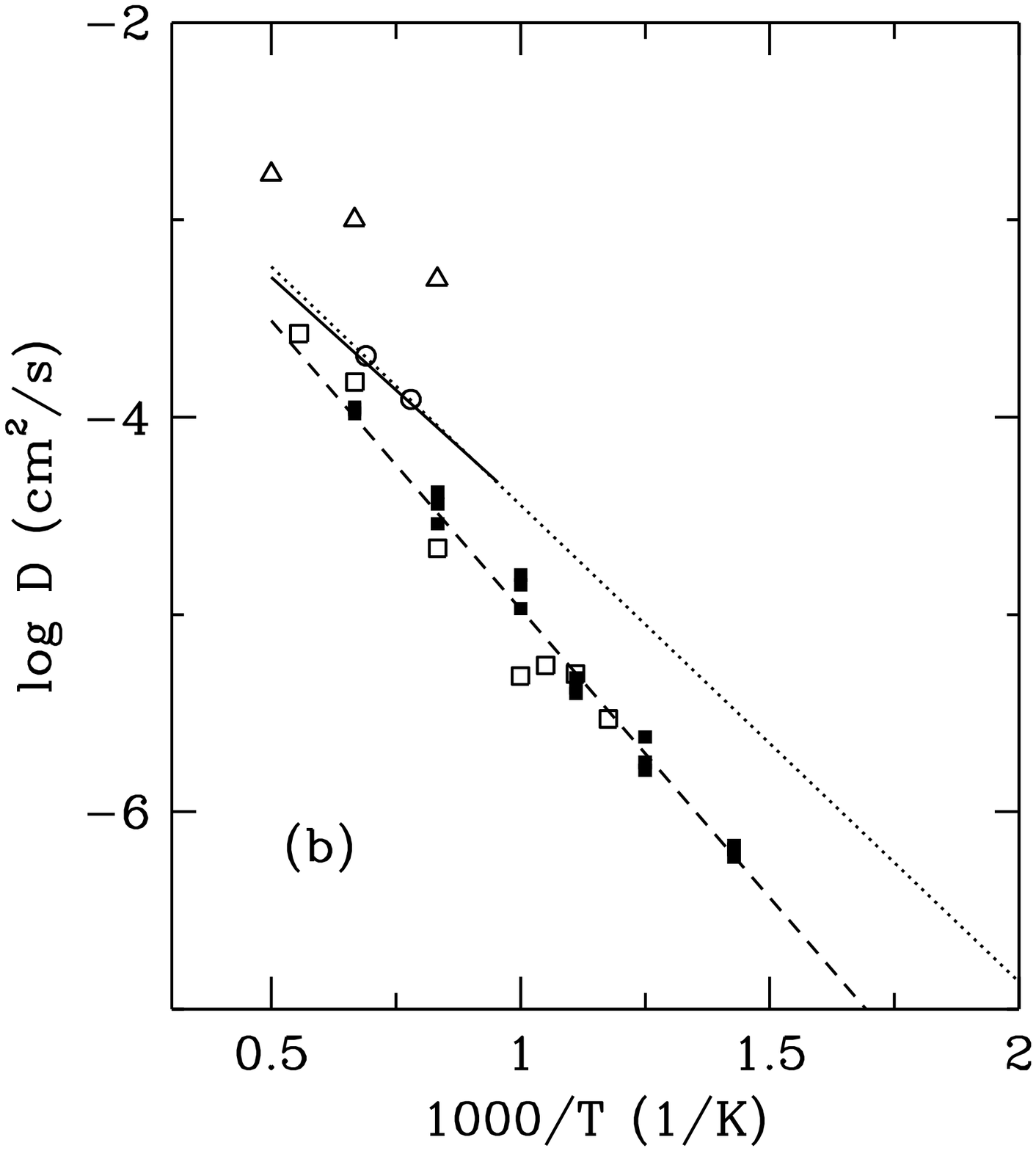}
\vspace{0in}
\caption{
(a) and (b) Coefficient of diffusion of hydrogen in crystalline silicon.
Panel (b) zooms in on the high temperature region. The present data are
indicated by the black squares; the dashed line is an Arrhenius fit to them.
Other data are as follows: open circles --- experimental data of Refs.\
\protect\onlinecite{vww56,ich68,sea88,mog85,tav88,pea85,joh92}; dotted line
--- fit to the high-temperature data of Ref.\ \protect\onlinecite{vww56} and
corresponding extrapolation to low temperatures; open squares --- TBMD data
of Panzarini and Colombo, Ref.\ \protect\onlinecite{pan94}; open triangles:
{\em ab initio} MD calculations of Buda {\em et al.}, Ref.\
\protect\onlinecite{bud89}; solid line: fit to the TBMD results of Boucher
and DeLeo, Ref.\ \protect\onlinecite{bou94}.
\label{diff}
}
\end{figure}

Experimental estimates of the diffusion constant are numerous and vary
widely,\cite{vww56,ich68,sea88,mog85,tav88,pea85,joh92} sometimes by two
orders of magnitude at a given temperature, as can be appreciated from the
open circles in Fig.\ \ref{diff}(a). This is an unpleasant state of affairs,
since precise knowledge of this quantity is important for both practical and
fundamental reasons: Because it forms complexes with a variety of defects,
hydrogen affects deeply the optical and electronic properties of
semiconductors. It is usually present as a result of the fabrication process,
but is often intentionally introduced in order to passivate defects. Being a
light species, further, H diffuses readily, inducing additional defects along
its way, thus affecting the transport properties of the material to an extent
which is determined by its relative concentration. It is therefore important
to understand diffusion at the atomic level in order to gain better control
on the properties of semiconductors and, in view of the simplicity of the
structure of the host material, it is of fundamental importance to be able to
understand this prototypical system.

The diffusion coefficient can be estimated, at sufficiently high temperature,
using the now well-established molecular-dynamics method; a proper model for
the interatomic potentials is then needed. Empirical potentials lack the
transferability and predictive power of first-principles methods. The latter,
however, are subject to limitations in size and time, which makes them
unpractical for the long simulations required for a proper (statistically
meaningful) estimate of the rate of diffusion. In an early application of the
Car-Parrinello method,\cite{car85} Buda {\em et al.} \cite{bud89} calculated
the coefficient of diffusion of H$^+$ in Si at three temperatures in the
range 1200--1950 K, covering a minuscule maximum observation time of 4 ps.

The semi-empirical, semi-quantum tight-binding molecular-dynamics (TBMD)
scheme, originally proposed by Khan and Broughton\cite{bro86} and Goodwin,
Skinner and Pettifor (GSP),\cite{goo89} provides good accuracy at a very
reasonable computational cost.\cite{lew98} Here, the attractive part of the
atom-atom interactions is described quantum-mechanically using (parametrized)
overlap (or hopping) integrals, while the repulsive part is fitted from known
properties of the system, e.g., binding energy vs distance. The forces are
then derived from the Hellman-Feynman theorem. TBMD models for Si:H were
proposed by Panzarini and Colombo (PC),\cite{pan94} Boucher and DeLeo
(BDL),\cite{bou94} and Kim, Lee and Lee; \cite{kim94} all three models are
based on the GSP model for Si--Si interactions, and use comparable fitting
schemes to take into account Si--H and H--H interactions.

The problem of hydrogen diffusion in silicon was addressed using TBMD by
Panzarini and Colombo \cite{pan94} as well as by Boucher and
DeLeo;\cite{bou94} both considered a single H atom in a 64-atom {\em c}-Si
supercell. While extending significantly the range of temperatures that were
covered by the {\em ab initio} simulations of Buda {\em et al.}\cite{bud89}
(1050--2000 K for BDL, 800--1800 K for PC), the timescales covered by these
MD calculations remain short (42 and 300 ps, respectively): indeed, for a
diffusion constant of $10^{-6}$ cm$^2$/s --- as found at about 800 K --- a
quick calculation indicates that the average size of the region visited by
the diffusing particle over 300 ps would be about 4 \AA, corresponding
roughly to the second-neighbour distance in {\em c}-Si. Further, the two
calculations exhibit some disagreement which might be inherent to the models
or, more likely, to the statistical quality of the MD data.

In this short note, we revisit the problem using, again, TBMD (PC version),
but with {\em much longer} timescales: our simulations ran during a
formidable 7 nanoseconds at the lowest temperature we could decently examine
--- 700 K, which is 100 K below the lowest temperature looked at by PC. Our
calculations generally confirm PC's results, in particular the discrepancies
with experiment observed at the lowest temperatures. However, while our
calculations indicate that long jumps dominate over single hops at high
temperatures, the diffusion coefficient exhibits no abrupt change with
temperature. We are led to conclude that the Arrhenius diffusion parameters
should ex\-tra\-po\-la\-te to low temperatures as far as the classical part
of the motion is concerned.

As previewed earlier, Fig.\ \ref{diff}(a) presents the results of several
measurements of the diffusion constant, plotted {\em \`a la} Arrhenius. Also
indicated are the {\em ab initio} MD data of Buda {\em et al.};\cite{bud89}
they are found to be in reasonable agreement with the high-temperature
experimental points of Van Wieringen and Warmoltz\cite{vww56}, fitted to the
Arrhenius law $D(T) = D_0\exp(-E_A/k_BT)$, with $D_0=9.41\times10^{-3}$
cm$^2$/s and $E_A=0.48$ eV. When extended to low temperatures [dotted line in
Fig.\ \ref{diff}(a)], one clearly sees the deviations from the Arrhenius
behaviour; it should be said, however, that there is no ``guarantee'' that
diffusion should be Arrhenius over the whole range of temperatures.

The TBMD data of PC and BDL are also plotted in Fig.\ \ref{diff}(a), and more
legibly in Fig.\ \ref{diff}(b). The agreement with experiment is clearly
excellent at high temperature --- certainly within the errors that can be
associated with both measurements and calculations. The data of BDL are found
to be extremely well fitted by the Arrhenius law with $D_0=6.91\times10^{-3}$
cm$^2$/s and $E_A=0.45$ eV all the way down to 1050 K, in striking agreement
with experiment, as can be judged by the close similarity between the
prefactors and energy barriers. PC, in contrast, observe significant
deviations from Arrhenius already at 1200 K, a problem that can possibly be
attributed to the statistical quality of their data.

\begin{figure}
\epsfxsize=3.0in \epsfbox{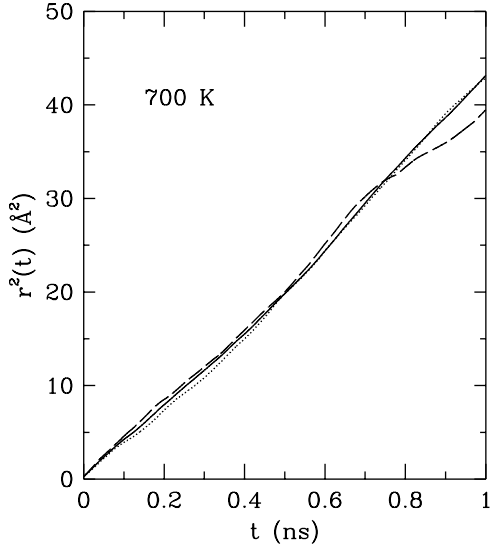}
\vspace{0in}
\caption{
Mean-square displacement of the hydrogen atom versus time at 700 K for a
maximum correlation time of 1 ns out of a 7 ns run. The three lines
correspond to averaging over the whole duration of the run (full line), the
first half (dotted line) and the second half (dashed line). The differences
between the three curves give an idea of the error bar on the diffusion
constant.
\label{r2}
}
\end{figure}

In Fig.\ \ref{r2} we give, as an example, the calculated time-dependence of
the mean-square displacement for our lowest-temperature run, viz.\ 700 K. The
simulation at this temperature ran for a total of 7 ns and the mean-square
displacement is calculated for a maximum correlation time of 1 ns in order to
minimize statistical uncertainties. We also evaluated the mean-square
displacement using only the first half of the run, then only the second half
so as to have a feeling for the ``error bar'' of our estimate of the
diffusion coefficient, $D = \lim_{t\rightarrow\infty} r^2(t)/6t$. The three
different calculations are indicated in Fig.\ \ref{r2} by full, dashed, and
dotted lines, respectively. The corresponding coefficients of diffusion,
which we discuss next, are presented in Fig.\ \ref{diff}; for each
temperature we investigated (700, 800, 900, 1000, 1200, and 1500 K), three
data points are thus given.

As can be appreciated from Fig.\ \ref{diff}(b), our estimates of the
diffusion coefficient generally agree with the values reported by PC, but
clearly with {\em much improved statistical quality}: the data points exhibit
very little scatter, falling along a single, rather well-defined straight
line. It is of course hazardous to assume that the diffusion process is
perfectly Arrhenius and that no ``exotic'' diffusion mechanisms (i.e., other
than single hops) are taking place. In order to ``guide the eye'', however,
we fitted our data to the law $D= 8.9 \times 10^{-3}\exp(0.58 {\rm eV}/k_BT)$
cm$^2$/s; this is displayed as the dashed line in Fig.\ \ref{diff}. Our
calculations, clearly, show no evidence of a sizeable change in the diffusive
behaviour as a function of temperature, as can possibly be (and was) inferred
from the data of PC.

Exotic mechanisms --- in the present case long jumps --- are in fact present;
this has already been noted by PC, who also found long jumps to be quenched in as
temperature drops; this observation was based on a visual examination of the
mean-square displacements. A physically more appropriate characterization of
single-atom motion is provided by the self part of the van Hove correlation
function (see, e.g., Ref.\ \onlinecite{han86}), $G_s({\bf r},t)$, which gives
the probability of finding a particle at ${\bf r}$ at time $t$ given that it
was at the origin at $t=0$. Averaging out over angular space, the function of
interest is $4 \pi r^2 G_s(r,t)$.

\begin{figure}
\epsfxsize=3.0in \epsfbox{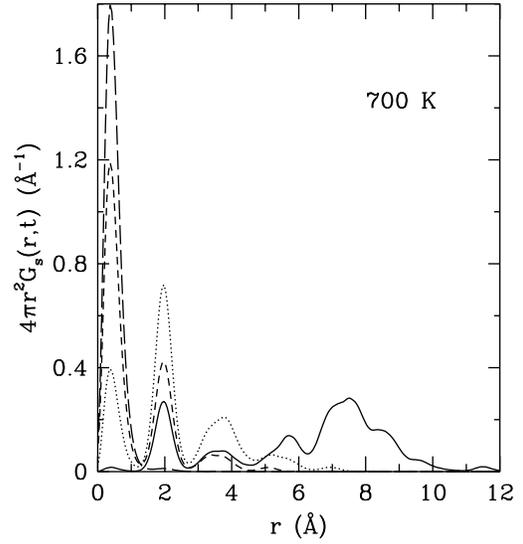}
\vspace{0in}
\caption{
Van Hove self-correlation function at 700 K as a function of distance for
four different correlation times: 1 ps (long dashes), 50 ps (short dashes),
200 ps (dots), and 1 ns (full line).
\label{vanhove}
}
\end{figure}

The van Hove self-correlation function for the H atom in {\em c-}Si at 700 K
is displayed in Fig.\ \ref{vanhove}. Here, $4 \pi r^2 G_s(r,t)$ is plotted as
a function of distance for four different times. At short time (but somewhat
longer than the typical vibrational period), single hops dominate the motion.
As time increases, the particle is (of course) on average further away and,
evidently, the diffusive motion proceeds via a sequence of single jumps ---
to a reasonable approximation.

Because diffusion is activated, it is more meaningful, in order to compare
the modes of diffusion at different temperatures, to examine the behaviour of
$4 \pi r^2 G_s(r,t)$ at {\em fixed} mean-square displacement. Thus, it is
possible in this way to determine the type of motion that leads a particle a
certain distance away from its original position. Here we chose (somewhat
arbitrarily) a fixed mean-square displacement of 50 \AA$^2$. From the
$r^2(t)$ curves, it is easy to determine the time at which, on average, and
for each temperature, the particle will be at the required fixed position.

\begin{figure}
\epsfxsize=3.0in \epsfbox{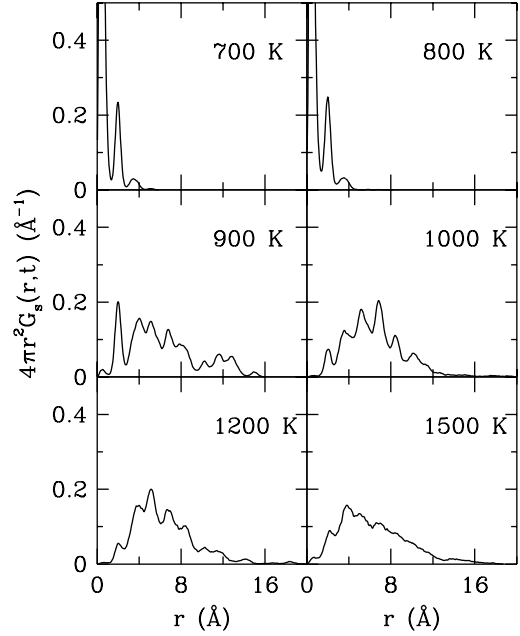}
\vspace{0.5in}
\caption{
Van Hove self-correlation function at various temperatures for {\em fixed}
average mean-square displacement (50 \AA$^2$), as explained in the text.
\label{tau}
}
\end{figure}

The results of this analysis are presented in Fig.\ \ref{tau}. Again, here,
we find that the motion consists of single hops at low temperatures, but
acquires a ``long-jump'' character as temperature increases. At the highest
temperatures, in fact, the distribution is nearly continuous, and single
jumps become almost undetectable. There is some evidence, in these plots, of
a discontinuous change in the manner that diffusion proceeds as a function of
temperature: upon going from 800 to 900 K, nearest-neighbour jumps are found
to become negligible, to the advantage of second- and further-neighbour hops.
But then no sign of such a change is apparent in the diffusion coefficient
(cf.\ Fig.\ \ref{diff}), which perhaps lacks the sensitivity needed to reveal
such details. Thus, we are led to conclude that the low-temperature ($T < $
about 900 K) diffusion coefficient is indeed Arrhenius-like, whereas
high-temperature diffusion proceeds via a complicated sequence of jumps whose
energy barriers combine into a single activation energy that is close to that
for single hops.

In view of this, there are no reasons to believe that this Arrhenius law
should not extrapolate to very low temperatures, {\em if the system were
classical}. This is clearly not the case here and it is of course expected
that diffusion will be enhanced by quantum contributions, even more so as
temperature decreases. We have not included quantum corrections in the
present study: this is a difficult problem and it is not clear that their
role is significant at the temperatures considered here. The small
discrepancies between TBMD and experimental data at high temperatures might
be related to quantum effects, but are perhaps more likely due to some
limitations of the model, i.e., are not significant. Likewise, Fig.\
\ref{diff} seems to indicate some partial ``recovery'' of the experimental
data at very low temperature. This might be a manifestation, again, of
quantum contributions. Clearly a more consistent set of experimental data is
needed in order to have a proper perspective on the problem.

{\it Aknowledgements ---}
We are grateful to Normand Mousseau for useful suggestions and to Edward
Hernandez for providing us with a TBMD code for the Goodwin-Skinner-Pettifor
Si model. We also thank Luciano Colombo for useful advice on implementing
the TBMD code for Si:H and for providing us an electronic version of Fig.\ 1
in Ref.\ \onlinecite{pan94}. This work is supported by grants from the
Natural Sciences and Engineering Research Council (NSERC) of Canada and the
``Fonds pour la formation de chercheurs et l'aide {\`a} la recherche'' (FCAR)
of the Province of Qu{\'e}bec.

\end{document}